\documentclass[12pt]{article}
\usepackage{amsmath,amssymb}
\usepackage{amsthm}
\usepackage{graphicx}
\usepackage{xcolor}
\usepackage[breaklinks=true,colorlinks,citecolor=teal,
linktocpage=true]{hyperref}
\usepackage{cite}
\usepackage[top=2.828cm,bottom=2.828cm, left=3cm,right=3cm]{geometry}
\usepackage{theorem}
\usepackage{epsf}
\usepackage{latexsym}
\usepackage{ascmac}

\newcommand{\del}{\partial}
\newcommand{\fr}{\frac}

\newcommand{\er}[1]{Eq.~\eqref{#1}}
\newcommand{\ers}[1]{Eqs.~\eqref{#1}}
\newcommand{\ket}[1]{|{#1}\rangle}
\newcommand{\bra}[1]{\langle{#1}|}

\newcommand{\vev}[1]{\langle #1 \rangle}
\newcommand{\hc}{\text{h.c.}}

\renewcommand{\b}{\bar}
\renewcommand{\d}{\dot}
\newcommand{\pd}[2]{\frac{\partial{#1}}{\partial{#2}}}

\newcommand{\sr}{\sqrt}
\newcommand{\bs}{\boldsymbol}
\newcommand{\df}{\dfrac}

\newcommand{\der}{\partial}

\renewcommand{\(}{\left(}
\renewcommand{\)}{\right)}

\newcommand{\dg}{\dagger}

\newcommand{\bmx}{\left(\begin{matrix}}
\newcommand{\emx}{\end{matrix}\right)}
\newcommand{\mtx}[1]{\bmx #1 \emx}
\newcommand{\non}{\nonumber\\}
\begin{document}
\pagenumbering{Roman}
\begin{titlepage}
\hfill December, 2018
\vspace{-1em}
\def\thefootnote{\fnsymbol{footnote}}%
   \def\@makefnmark{\hbox
       to\z@{$\m@th^{\@thefnmark}$\hss}}%
 \vspace{3em}
 \begin{center}%
  {\Large\bf 
Supersymmetry breaking and ghost Goldstino
\\
 in modulated vacua
  \par
   }%
 \vspace{1.5em}
{
Sven Bjarke Gudnason,${}^1$\mbox{}\footnote{gudnason(at)keio.jp}
Muneto Nitta,${}^1$\mbox{}\footnote{nitta(at)phys-h.keio.ac.jp}
Shin Sasaki,${}^2$\mbox{}\footnote{shin-s(at)kitasato-u.ac.jp}
and
Ryo Yokokura${}^1$\mbox{}\footnote{ryokokur(at)keio.jp}
   \par
}%
 \vspace{1.5em}
{\small \it
${}^1$Department of Physics, and Research and Education Center for Natural Sciences, \\
Keio University, Hiyoshi 4-1-1, Yokohama, Kanagawa 223-8521, Japan
 \vspace{0.5em}
\\
${}^2$Department of Physics, Kitasato University, Sagamihara 252-0373, Japan
 \par
}
\vspace{1.5em} 
   {\large 
    }%
 \end{center}%
 \par
\vspace{1.5em}%
\begin{abstract}
We discuss spontaneous supersymmetry (SUSY) breaking mechanisms by
means of modulated vacua in four-dimensional ${\cal N} =1$
supersymmetric field theories. 
The SUSY breaking due to spatially modulated vacua is
extended to the cases of temporally and lightlike modulated vacua,
using a higher-derivative model with a chiral superfield, free from
the Ostrogradsky instability and the auxiliary field problem.
For all the kinds of modulated vacua, SUSY is spontaneously
broken and the fermion in the chiral superfield becomes a Goldstino.
We further investigate the stability of the modulated vacua.
The vacua are (meta)stable if the vacuum energy density is 
non-negative. 
However, the vacua become unstable due to the presence of the ghost
Goldstino if the vacuum energy density is negative.
Finally, we derive the relation between the presence of the ghost
Goldstino and the negative vacuum energy density in the modulated
vacua using the SUSY algebra. 
\end{abstract}
\end{titlepage}
\pagenumbering{arabic}
\setcounter{footnote}{0}%
\def\thefootnote{\arabic{footnote}}%
   \def\@makefnmark{\hbox
     to\z@{$\m@th^{\@thefnmark}$\hss}}%
\tableofcontents
\section{Introduction}

Understanding the vacuum structure of the quantum field theory under
study is the starting point for any analysis. Toy models may
have a simple vacuum structure in which the fields are energetically
preferred to sit at the origin of the field space, retaining the
symmetries present in the Lagrangian formulation.

Another example of vacua is that of QCD, which is expected to be far
from trivial and dynamically generate a mass gap giving mass to the
lightest glue state---the glueball---and to the hadrons of the
theory when coupled to fermions.

In a recent series of papers, we have studied nontrivial vacua in
which the vacuum expectation value (VEV) is not a constant but has a
phase that winds along a spatial direction~\cite{Nitta:2017mgk,Nitta:2017yuf,Gudnason:2018bqb}, 
along a temporal direction~\cite{Gudnason:2018bqb} 
or along the direction of the light cone~\cite{Gudnason:2018bqb}. 
Conceptually, we can think of this construction as an intermediate 
situation between the trivial vacua with a vanishing VEV and the
enormously complicated vacuum of QCD.
Our construction is inspired by the so-called Fulde-Ferrel (FF) 
state~\cite{Fulde:1964zz}, which is the lowest-energy state in certain
condensed matter systems, such as a superconducting ring with a magnetic
field applied perpendicularly \cite{Yoshii:2014fwa} (for a review see,
e.g., Ref.~\cite{FFLO-review}). 
The FF state also exists in QCD itself (Nambu-Jona--Lasino model)  
at finite density and temperature and is called a dual chiral density
wave or a chiral spiral \cite{Nakano:2004cd,Karasawa:2013zsa}. 
In these cases, Lorentz invariance is absent due to the finite
temperature and/or density. 
In contrast, our construction works in a Lorentz invariant theory at
vanishing density; however, it relies on the use of higher-derivative
operators and the imposition of shift symmetry.
In Ref.~\cite{Nitta:2017yuf}, we have shown that the global stability
of this class of model dictates that the highest-derivative term must
have $2(2\ell+1)$ derivatives with $\ell\in\mathbb{Z}_{>0}$. 
The simplest class of models that contains a phase-modulated (FF-type)
vacuum in the spatial, temporal, and lightlike directions has a
sixth-order derivative term as the highest-derivative term.
This model interestingly has, as a submodel, a supersymmetric extension 
\cite{Nitta:2017yuf}.

It is well known that terms in the Lagrangian with more than one
spacetime derivatives on one field, cause an instability
of the system. This is called the Ostrogradsky instability
\cite{Ostrogradski} which substantially results in ghost states in the
quantum regime.
At the classical level, it implies the loss of a lowest-energy state. 
We therefore focus on models where only higher-dimension operators
given by the first-order derivatives of fields appear such as
$(\der\varphi)^{2n}$ for a scalar field $\varphi$.

Higher-derivative terms in supersymmetric field theories are quite
nontrivial as they generically suffer from a problem called the
auxiliary field problem.
More precisely, in generic supersymmetric higher-derivative models,
the equation of motion (EOM) for the auxiliary field $F$ is not
necessarily algebraic \cite{Gates:1995fx,Gates:1996cq}. The result is
that eliminating $F$ and finding the on-shell Lagrangian is essentially
impossible. 
Usually, this auxiliary-field problem comes with an Ostrogradsky ghost
\cite{Antoniadis:2007xc,Dudas:2015vka}, but there is also an exception
\cite{Fujimori:2016udq,Fujimori:2017rcc}. 

We therefore look for supersymmetry (SUSY) models 
that do not suffer from the
Ostrogradsky instability nor from the auxiliary field problem.
A natural candidate for such a model is the higher-derivative chiral
SUSY model studied in
Refs.~\cite{Khoury:2010gb,Koehn:2012ar,Koehn:2012te,Adam:2013awa,Nitta:2014pwa,Nitta:2014fca,Nitta:2015uba,Nitta:2017yuf}.
This latter model canonically gives a supersymmetric fourth-order
derivative term, multiplied by a function of the superfield,
$\Lambda(\Phi)$. Because of the fourth-order term being saturated in the
nilpotent series of Grassmann numbers, the Grassmannian integral only
picks up the bosonic component of the function $\Lambda$ and hence it
is straightforward to construct a sixth-order derivative model this
way.
The model constructed this way, turns out to be exactly a submodel of
the phase-modulated higher-derivative scalar field theory models that
we constructed in Refs.~\cite{Nitta:2017mgk,Gudnason:2018bqb}.
One of the interesting features is that SUSY is spontaneously broken
due to derivatives of the field, $\partial\varphi$, in contrast to the
conventional cases in which a nonzero SUSY auxiliary field,  
$F$ term or $D$ term, breaks SUSY.

In this paper, we study the SUSY breaking in all kinds of the
phase-modulated (FF-type) vacua, i.e.~spatially, temporally,
and lightlike modulated vacua. 
First, we will review the construction of the phase-modulated 
vacuum solutions of
Refs.~\cite{Nitta:2017mgk,Nitta:2017yuf,Gudnason:2018bqb} in the 
cases of spatially, temporally, and lightlike modulated vacua.
Then we will discuss the fluctuations about these modulated vacuum
solutions, for both the scalar field and the fermion.
The main new result in this paper is that we derive a relation between
the Goldstino and the vacuum energy density of the vacua in the
models.
After the discussion of the modulated vacua and the ghost Goldstinos  
in a concrete model, we rederive the same
result in a model-independent way by using only the SUSY algebra and
the knowledge of the broken/unbroken symmetries of the vacuum.

The plan of the paper is as follows.
In Sec.~\ref{sec:SUSYmodel} we review the type of higher-derivative
chiral SUSY model that is a unique candidate for avoiding the
Ostrogradsky problem and auxiliary field problem and lies in the class
of models that can possess modulated vacua of
Refs.~\cite{Nitta:2017mgk,Nitta:2017yuf,Gudnason:2018bqb} 
whose construction we
review in Sec.~\ref{sec:modvac}.
Sec.~\ref{sec:fluc} reviews the bosonic fluctuation spectra and
introduces the main new result, which comes from studying the
fermionic fluctuations and finding the relation to the vacuum energy
density.
This latter relation is then studied using the SUSY algebra in
Sec.~\ref{gg}.
Finally, Sec.~\ref{sec:summary} concludes with a summary and a
discussion of the open problems.

\section{Higher-derivative SUSY model}\label{sec:SUSYmodel}
In this section, we introduce a supersymmetric model in which
modulated vacua of the FF-type is allowed. 
Since modulated vacua are characterized by a nonzero 
VEV of spacetime derivatives of a scalar field
$\del_m\varphi$, it is necessary to introduce a ``potential'' of the
derivative term $\del_m\varphi$ for it to develop a nonzero
VEV.\footnote{A similar mechanism for nonzero VEVs of $\del_m\varphi$
  is discussed in the context of ghost condensation
  \cite{ArkaniHamed:2003uy}, in which case, due to the wrong sign of
  the canonical kinetic term, $\dot{\varphi}$ develops a nonzero
  VEV.} 
This inevitably results in models where the bosonic part of the
Lagrangian consists of terms with polynomials of $\del_m\varphi$,
i.e., higher-derivative SUSY models. 
In order to consider higher-derivative models, it is convenient to
work in the off-shell superfield formalism.
The four-dimensional $\mathcal{N} = 1$ superspace is characterized by the
bosonic spacetime coordinates $(x^m)$ ($m = 0,1,2,3$) and the
fermionic coordinates given by Grassmann numbers  
$(\theta^\alpha,\b\theta_{\d\alpha})$.
Here, the Greek letters beginning with $\alpha,\beta,\ldots$
and $\d\alpha, \d\beta,\ldots$ denote undotted and dotted spinors, 
respectively.
We use the notation and conventions of Ref.~\cite{Wess:1992cp}
throughout this paper.

We introduce a chiral superfield, $\Phi$, which contains a complex
scalar field $\varphi$.
This is utilized to describe VEVs in modulated vacua.
The component fields of the chiral superfield, $\Phi$, are defined as 
\begin{equation}
\varphi = \Phi|, 
\quad
\psi_\alpha = \fr{1}{\sr{2}} D_\alpha \Phi|,
\quad
F = - \fr{1}{4} D^2 \Phi|,
\end{equation}
where $D_{\alpha}$ is the supercovariant derivative and 
the symbol $|$ denotes that the values are evaluated at
$\theta^{\alpha} = \bar{\theta}_{\dot{\alpha}} = 0$. 
The field $\psi_{\alpha}$ is a Weyl fermion and $F$ is an auxiliary
field. 

We now discuss the supersymmetric higher-derivative chiral models.
As already mentioned in the Introduction, we here focus on models
that only depend on the first derivative of the fields. This will
sidestep the issue of the Ostrogradsky instability.
Furthermore, to avoid the auxiliary field problem, we work in the
higher-derivative chiral SUSY model of
Refs.~\cite{Khoury:2010gb,Koehn:2012ar,Koehn:2012te,Adam:2013awa,Nitta:2014pwa,Nitta:2014fca,Nitta:2015uba,Nitta:2017yuf,Nitta:2018yzb,Nitta:2018vyc},
which is known to be free from this problem as the EOM for $F$ remains
algebraic. 
The Lagrangian is given by \footnote{
  The case of a constant $\Lambda$ was first found long ago 
  \cite{Buchbinder:1994iw,Buchbinder:1994xq}, 
  the dependence of $\Lambda$ on $\Phi,\b\Phi$ was introduced in 
  Refs.~\cite{Khoury:2010gb,Koehn:2012ar,Koehn:2012te,Adam:2013awa},
  the dependence on $\der_m\Phi, \der_m\b\Phi$ was found 
  in Refs.~\cite{Nitta:2014pwa,Nitta:2014fca,Nitta:2015uba},
  and finally the dependence on 
  $D^2 \Phi,\b{D}^2 \b\Phi$ was found in Refs.~\cite{Nitta:2018yzb,Nitta:2018vyc}.
} 
\begin{align}
{\cal L}
&= 
\int d^4\theta\; K(\Phi,\b\Phi) 
+\(\int d^2\theta\; W(\Phi) + \hc\)
\non
&\phantom{=\ }
+
\fr{1}{16}\int d^4\theta\; \Lambda(\Phi, \Phi, \der_m\Phi, \der_m\b\Phi, D^2 \Phi,\b{D}^2 \b\Phi)
(D^\alpha \Phi)(D_\alpha \Phi) (\b{D}_{\d\alpha}\b\Phi)
(\b{D}^{\d\alpha}\b\Phi). 
\label{181108.0249}
\end{align}
Here, $K$, $W$ and $\Lambda$ are a K\"ahler potential, a
superpotential and a real scalar function, respectively.
The right-hand side of the first line gives us a quadratic kinetic
term and a potential term for $\varphi$, while the second line leads
to higher-derivative terms. 

As discussed in Ref.~\cite{Gudnason:2018bqb}, in order to realize
modulated vacua in our construction, it is necessary to introduce at
least sixth-order derivative terms $(\del\varphi)^{6}$.
This is based on the global stability of the vacua.
For simplicity, we also assume that the model possesses shift symmetry
$\varphi\to\varphi+c$ where $c$ is a complex constant. 
The simplest model
that accommodates these conditions is 
\begin{equation}
  K = k \Phi\b\Phi, \quad
  W = 0, \quad
  \Lambda = \lambda  + \alpha \der^m \Phi \der_m \b\Phi,
\end{equation}
where $k$, $\lambda$ and $\alpha$ are real constants.
Therefore, the model is given by 
\begin{equation}
  {\cal L} = \int d^4\theta\; k\Phi \b\Phi
  +\fr{1}{16}\int d^4\theta\; (\lambda + \alpha \der^m \Phi \der_m\b\Phi)
(D^\alpha \Phi)(D_\alpha \Phi) (\b{D}_{\d\alpha}\b\Phi)
(\b{D}^{\d\alpha}\b\Phi).
\label{181108.0206}
\end{equation}
The above model was proposed in Ref.~\cite{Nitta:2017yuf} where SUSY
breaking in a spatially modulated vacuum is discussed. 

In the Lagrangian, there is an auxiliary field $F$ that does not
have physical degrees of freedom. 
We eliminate the auxiliary field by the EOM.
In order to obtain the EOM, it is convenient to write out the component
Lagrangian from \er{181108.0206}. 
The bosonic part of the Lagrangian is 
\begin{align}
 {\cal L}_{\rm boson}
&= -k \der^m \varphi \der_m \b\varphi
 + k F\b{F}
\non
&\phantom{=\ } 
+(\lambda  + \alpha \der^m\varphi\der_m\b\varphi)
\(
 (\der^n \varphi \der_n\varphi)
(\der^p\b\varphi\der_p\b\varphi)
-2 F\b{F} \der^n \varphi \der_n\b\varphi + F^2 \b{F}^2
\),
\end{align}
where we have omitted the fermions, since the fermions will be irrelevant 
to find the modulated vacua.
Note that the fermionic part will be used when we discuss the
fluctuation of the Goldstino.
The EOM for the auxiliary field is 
\begin{equation}
  k F
  -2F (\lambda + 2\alpha \der^m\varphi \der_m\b\varphi)
  \(\der^n\varphi\der_n\b\varphi - |F|^2\) = 0.
\end{equation}
As advertised above, the equation is algebraic {\it i.e}.~it does not
involve terms with spacetime derivatives of $F$.
We can therefore easily find solutions to this equation and they are
\begin{equation}
 F = 0, \quad |F|^2 = 
-\fr{k}{2(\lambda +\alpha \der^m\varphi\der_m \b\varphi)}
+ \der^m \varphi \der_m \b\varphi.
\end{equation}
We note that these are exact analytic solutions in the bosonic sector,
but including fermions is somewhat cumbersome.
They can be incorporated in the solutions perturbatively as we will
see in Sec.~\ref{sec:fluc} (see also Ref.~\cite{Nitta:2017yuf} for the
detailed analysis). 
There are two distinct on-shell branches corresponding to these
solutions.
For the first solution, the on-shell Lagrangian is  
\begin{equation}
 {\cal L}_{\rm boson} = 
 -k \der^m \varphi \der_m \b\varphi
+\lambda 
(\der^m \varphi \der_m\varphi)  (\der^n\b\varphi\der_n\b\varphi)
+\alpha (\der^m \varphi \der_m \b\varphi) (\der^n \varphi \der_n\varphi)
  (\der^p\b\varphi\der_p\b\varphi).
\label{181108.0330}
\end{equation}
Thus, the model contains a fourth- and a sixth-order derivative of the 
complex scalar field $\varphi$, as expected.
On this branch, the higher-derivative terms are introduced
perturbatively in addition to the canonical (quadratic) kinetic term. 
This is the so-called canonical branch.
Several supersymmetric higher-derivative models are constructed using
this branch including Dirac-Born-Infeld models
\cite{Rocek:1997hi,Sasaki:2012ka}, supersymmetric 
$P(X,\phi)$ models
\cite{Khoury:2010gb,Koehn:2012ar,Koehn:2012te}, 
higher-derivative corrections to a low-energy
effective theory \cite{Nitta:2014fca}, and so on.

For the second solution in \er{181108.0330}, the on-shell Lagrangian is  
\begin{equation}
{\cal L}_{\rm boson,nc} = 
\(|\der_m\varphi\der^m\varphi|^2 
-(\der_m \varphi\der^m\b\varphi)^2\)
(\lambda +\alpha \der^n\varphi \der_n\b\varphi)
-\fr{k^2}{\lambda +\alpha \der^m\varphi \der_m\b\varphi}.
\label{181108.2213}
\end{equation}
In this Lagrangian, the canonical (quadratic) kinetic term 
vanishes.
This branch is the so-called noncanonical
branch~\cite{Khoury:2010gb,Koehn:2012te,Nitta:2014pwa}.
On this branch, the higher-derivative terms are not introduced
perturbatively because we cannot take the limit $\lambda\to 0$ or
$\alpha\to 0$.
On the noncanonical branch, supersymmetric (baby-)Skyrme models have
been discussed in
Refs.~\cite{Adam:2013awa,Bolognesi:2014ova,Gudnason:2015ryh,Queiruga:2015xka,Gudnason:2016iex}. 
Since the model that allows modulated vacua has a quadratic kinetic
term, we use the first solution and its on-shell Lagrangian in
\er{181108.0330} rather than that of \er{181108.2213}.

\section{Modulated vacua in SUSY theories}\label{sec:modvac}

In this section, we examine modulated vacua in the model
\eqref{181108.0330}.  
First, we will discuss the general arguments for the modulated vacua.
Second, we find the modulated vacua as solutions to the EOMs and
energy-extremum conditions. 
We also calculate the energy density in the modulated vacua, which
will be used in the later discussion.
This section is mostly a review of the results in
Ref.~\cite{Gudnason:2018bqb}. 
In the following, we give a brief overview of the general discussion
of modulated  vacua in Lorentz-invariant field theories. 
For more details, see Ref.~\cite{Gudnason:2018bqb}.

\subsection{General discussion}
In the ordinary situation where VEVs are constants, they are
determined by the extremal condition of the energy density. 
In this case, the VEVs solve the EOM automatically.
On the other hand, the latter condition is not trivial for modulated
vacua since in that case the VEVs depend on spacetime coordinates.
In the following, we will write down the conditions for modulated
vacua to solve both the energy-extremal condition and the EOM. 

To find vacua, we solve the EOM and the energy-extremum condition
for the complex scalar field $\varphi$ using the Ansatz 
$\vev{\psi_\alpha}=\vev{\b\psi_{\d\alpha}}=0$.
The EOM for $\varphi$ is generically
\begin{equation}
  0 = \der_m \pd{{\cal L}_{\rm boson}}{\der_m \varphi}
  = \pd{^2 {\cal L}_{\rm boson}}{(\der_m \varphi)\der (\der_n \varphi)}
  \der_m\der_n\varphi +
  \pd{^2 {\cal L}_{\rm boson}}{(\der_m \varphi)\der (\der_n \b\varphi)}
  \der_m\der_n\b\varphi, 
\label{181001.2040}
\end{equation}
and the EOM for $\b\varphi$ is the complex conjugate of the above
equation. 
The EOM together with its complex conjugate can be written in matrix
form as follows:
\begin{equation}
0=
 {\bf L}^{mn} 
\mtx{
\der_m \der_n \varphi
\\
\der_m \der_n \b\varphi
}
=
{\bf L}^{00} 
\mtx{\ddot\varphi \\ \ddot{\b\varphi}}
+
({\bf L}^{0i} + {\bf L}^{i0})
\mtx{
\der_i \d\varphi
\\
\der_i \d{\b\varphi}
}
+
{\bf L}^{ij}
\mtx{
\der_i\der_j \varphi
\\
\der_i\der_j \b\varphi}.
\label{181120.2101}
\end{equation}
Here, ${\bf L}^{mn}$ is defined as
\begin{equation}
 {\bf L}^{mn} 
:= 
\mtx{
\pd{^2 {\cal L}}{(\der_m\b\varphi)\der(\der_n \varphi)}
&
\pd{^2 {\cal L}}{(\der_m\b\varphi)\der(\der_n \b\varphi)}
\\
\pd{^2 {\cal L}}{(\der_m\varphi)\der(\der_n \varphi)}
&
\pd{^2 {\cal L}}{(\der_m\varphi)\der(\der_n \b\varphi)}
}.
\end{equation}
In \er{181120.2101}, 
we have split the spacetime derivative $\der_m\varphi $ 
into the temporal direction $\der_0 \varphi = \d\varphi$ and 
the spatial directions $\der_i \varphi $ ($i = 1,2,3$),
since we will discuss the 
temporally, spatially, and lightlike modulated vacua.
Vacua in field theories are characterized by 
(local) minima of the energy functional.
The energy density (Hamiltonian) is defined as 
\begin{equation}
 {\cal H} := 
\pd{{\cal L}}{\d\varphi} \d\varphi 
+
\pd{{\cal L}}{\d{\b\varphi}} \d{\b\varphi} -{\cal L}.
\label{181128.2203}
\end{equation}
Since the energy density is written in terms of $\der_m \varphi$ 
and its conjugate, the minima of the energy satisfy
the following conditions:
\begin{equation}
  0= \pd{{\cal H}}{\der_m \varphi} 
\end{equation}
and its complex conjugate.
The conditions can be rewritten in terms of ${\bf L}^{mn}$ as
\begin{equation}
0= {\bf L}^{00} \mtx{\d\varphi \\ \d{\b\varphi}},
\quad
0= {\bf L}^{i0} \mtx{\d\varphi \\ \d{\b\varphi}} 
- 
\mtx{\pd{{\cal L}}{\der_i \b\varphi} \\ \pd{{\cal L}}{\der_i \varphi}}.
\label{181120.2128}
\end{equation}
The modulated vacua are characterized by the solutions to 
\ers{181120.2101} and \eqref{181120.2128}.

In the vacua, spacetime or internal symmetries are generally
broken. 
The vacua can be classified by the broken translational generators
$P^{\hat{m}}$. 
If $P^{\hat{m}}$ is spacelike, timelike, or lightlike (null), the
vacua are called spatially, temporally or lightlike modulated vacua,
respectively.
In Ref.~\cite{Gudnason:2018bqb}, the conditions for the presence of
the spatially, temporally, or lightlike modulated vacua are studied
systematically. 
Since parts of the symmetries in the theory are broken, it is natural
to study the Nambu-Goldstone (NG) modes in the vacuum.
In the ordinary cases where VEVs are constants, the NG modes
correspond to the flat directions of the potential term. 
The zero modes of the Hessian matrix associated with the curvature of 
the potential correspond to the NG modes. This implies that the quadratic
term of the NG modes---the mass term---vanishes, and hence they are
massless modes. 
In our setup, however, there are no ordinary potential terms but
instead a ``potential'' for the derivatives of the fields. 
Therefore, we found that it is useful to consider the notion of the
\emph{generalized Nambu-Goldstone modes} to examine the flat
directions in the modulated vacuum
\cite{Nitta:2017mgk,Gudnason:2018bqb}. 
Similar to ordinary NG modes, the generalized NG modes correspond to 
zero modes of the Hessian matrix (the generalized mass matrix):
\begin{align}
\mathcal{M} =
\left(
\begin{array}{cccc}
\mathbf{M}^{00} & \mathbf{M}^{01} & \cdots & \mathbf{M}^{03}
 \\
\mathbf{M}^{10} & \mathbf{M}^{11} & \cdots & 
\\
\vdots & & \ddots & 
\\
\mathbf{M}^{30} & \cdots & & \mathbf{M}^{33} 
\end{array}
\right),
\label{eq:generalized_mass_matrix}
\end{align}
where $\mathbf{M}^{mn}$ are given by
\begin{align}
\mathbf{M}^{mn} =
\left(
\begin{array}{cc}
\frac{\del^2 \mathcal{H}}{\del (\del_m \bar{\varphi}) \del (\del_n
 \varphi)} & \frac{\del^2 \mathcal{H}}{\del (\del_m \bar{\varphi})
 \del (\del_n \bar{\varphi})} 
 \\
\frac{\del^2 \mathcal{H}}{\del (\del_m \varphi) \del (\del_n \varphi)} & 
\frac{\del^2 \mathcal{H}}{\del (\del_m \varphi) \del (\del_n \bar{\varphi})}
\end{array}
\right).
\label{eq:Mmn_def}
\end{align}
Now that we have the general conditions and material to analyze
modulated vacua, we will in the next subsection solve the conditions
\ers{181120.2101} and \eqref{181120.2128} in the cases of temporally,
spatially, or lightlike modulated vacua.

\subsection{Spatially modulated vacua}
First, we consider the spatially modulated vacua.
We employ the following Ansatz for solving \ers{181120.2101} and
\eqref{181120.2128}: 
\begin{equation}
\vev{\varphi}  = \varphi_0 e^{i c  x^1},
\quad
\vev{\d\varphi} = \vev{\der_{2,3} \varphi} =
\vev{ \psi_\alpha} 
= \vev{\b\psi_{\d\alpha}} 
= \vev{F} = \vev{\b{F}} =0.
\label{181121.0217}
\end{equation}
Here, $\varphi_0$ and $c$ are complex and real constants,
respectively.
Within the Ansatz, the energy-extremum condition in \er{181120.2128}
becomes 
\begin{equation}
 0 = \der_1 \b\varphi 
  \(-k +2\lambda |\der_1\varphi|^2 +3 \alpha |\der_1\varphi|^4\).
\end{equation}
When $\lambda^2+3\alpha k>0$, there is a local minimum in the energy
potential for which $\varphi_0$ is nonzero. 
In this case, the above condition determines the amplitude of the VEV: 
\begin{equation}
 |\der_1\varphi|^2 =
 c^2 |\varphi_0|^2 =
 \fr{-\lambda \pm \sr{\lambda^2 +3\alpha k}}{3\alpha}.
\end{equation}
Since $|\der_1\varphi|^2$ is positive, the parameter $\alpha$ should
be negative. 
As discussed in Ref.~\cite{Gudnason:2018bqb}, for the Ansatz
\eqref{181121.0217}, we have the relation $\mathcal{H}=-\mathcal{L}$
and this implies that the extremal condition of $\mathcal{H}$ is
equivalent to the EOM given by the first variation of
the Lagrangian: $\delta\mathcal{L}=0$. 
Therefore, solutions to \er{181120.2101} automatically satisfy the
condition \eqref{181120.2128}.
Indeed, in the Ansatz \eqref{181121.0217}, the right-hand side of the
EOM in \er{181120.2101} reduces to
\begin{equation}
 {\bf L}^{11} \mtx{\der^2_1 \varphi \\ \der_1^2\b\varphi} = ic 
\mtx{
\der_1 \varphi\(-k 
+ 2\lambda |\der_1\varphi|^2 + 3\alpha |\der_1\varphi|^4\) \\
-\der_1 \b\varphi\(-k 
+ 2\lambda |\der_1\varphi|^2 + 3\alpha |\der_1\varphi|^4\)},
\end{equation}
which automatically vanishes if the energy-extremum condition is
satisfied.
In this vacuum, the VEV of the vacuum-energy density 
${\cal E}_{\rm sp} $ is 
\begin{align}
{\cal E}_{\rm sp} &= k |\der_1 \varphi |^2 - \lambda |\der_1 \varphi |^4
- \alpha |\der_1 \varphi|^6 \non
&= -\fr{1}{27\alpha^2}
  \(\lambda +\sr{\lambda^2 + 3\alpha k}\)
  \(6\alpha k +\lambda (\lambda +\sr{\lambda^2 + 3\alpha k})\).
\label{181011.1320}
\end{align}
The energy density can be positive, zero or negative depending on 
the parameters.
If $\lambda^2 < -4\alpha k$, the energy density is positive.
In this case, the modulated vacuum is metastable.
If $\lambda^2 = -4\alpha k$, the energy density is zero.
In this case, the trivial vacuum $\der_1\varphi=0$ and the modulated
vacuum has the same energy density. 
If $\lambda^2 > -4\alpha k$, the energy density is negative.
In this case, the modulated vacuum is energetically favored.

Finally, we discuss the spontaneous breaking of the symmetries in the
spatially modulated vacuum. 
Because of the nonzero VEV of $\varphi=\varphi_0 e^{icx^1}$, the
translational transformation $P^{1}$, the global $U(1)$
transformation, shift transformation $S$, and the Lorentz
transformations $M^{1m}$ become broken generators.
However, since the simultaneous transformation of $P^1$ and 
the global $U(1)$ transformation is preserved, the symmetry breaking
pattern is  
$ISO(3,1)\times U(1)\times S\to ISO(2,1)\times [U(1)\times {\cal P}^1]_{\rm diag.}$.
Here, $ISO$ denotes the Poincar\'e group, and ${\cal P}^m $ denotes
the spacetime translational group.
We also note that, in the vacuum \eqref{181121.0217}, the SUSY
variation of the fermion is nonvanishing: 
\begin{align}
\delta \psi_{\alpha} = i \sqrt{2} (\sigma^m)_{\alpha \dot{\alpha}}
 \bar{\xi}^{\dot{\alpha}} \partial_m \varphi + \sqrt{2} \xi_{\alpha} F =
 i \sqrt{2} (\bar{\sigma}^1)_{\alpha \dot{\alpha}}
 \bar{\xi}^{\dot{\alpha}} \partial_1 \varphi.
\label{eq:fermion_variation}
\end{align}
Here $\xi,\bar{\xi}$ are SUSY transformation parameters. 
Therefore, SUSY is spontaneously broken in the spatially 
modulated vacuum. We note that the condition
\eqref{eq:fermion_variation} holds for any values of the energy 
density. 
Namely, SUSY can be broken even for zero vacuum energy density in
supersymmetric higher-derivative models.
This is in contradistinction to the ordinary situation in which vacua
are given by the extrema of potentials and VEVs are constants.

\subsection{Temporally modulated vacua}\label{sec:tmv}
For the temporally modulated vacua, we will solve \ers{181120.2101}
and \eqref{181120.2128} by using the following Ansatz:
\begin{equation}
\vev{\varphi}  = \varphi_0 e^{i \omega x^0}, \quad
 \vev{\der_i \varphi} =\vev{ \psi_\alpha} 
= \vev{\b\psi_{\d\alpha}} 
= \vev{F} = \vev{\b{F}} = 0.
\label{181121.1258}
\end{equation}
Here, $\varphi_0$ and $\omega$ are complex and real constants,
respectively. 
By the Ansatz, 
the condition for the extremum of the energy in
\er{181120.2128} is reduced to 
\begin{equation}
{\bf L}^{00} \mtx{\d\varphi \\ \d{\b\varphi}} = 0, 
\end{equation}
while the EOM in \er{181001.2040} is
\begin{equation}
 {\bf L}^{00} \mtx{\ddot\varphi \\ \ddot{\b\varphi}} = 0.
\end{equation}
Here, the matrix ${\bf L}^{00}$ is calculated as
\begin{equation}
 {\bf L}^{00}
 = 
\mtx{k +4\lambda |\d\varphi|^2 -9\alpha |\varphi|^4
& 2\d\varphi^2 (-3\alpha |\d\varphi|^2 +\lambda) \\
 2\d{\b\varphi}^2 (-3\alpha |\d\varphi|^2 +\lambda)
&
k +4\lambda |\d\varphi|^2 -9\alpha |\varphi|^4}.
\end{equation}
These conditions are satisfied by ${\bf L}^{00}=0$.
The conditions lead to the temporally modulated vacua
\begin{equation}
|\vev{\d\varphi}|^2 = \fr{\lambda}{3\alpha},
\label{181001.2100}
\end{equation}
with the condition on the parameter
\begin{equation}
k = -\fr{\lambda^2}{3\alpha}.
\label{181121.1257}
\end{equation}
If we further assume $k > 0$, both of the parameters $\lambda$ and
$\alpha$ must be negative.

Now we calculate the vacuum energy of the temporally modulated vacua. 
Since the parameters are restricted by \er{181121.1257}, in contrast
to the spatially modulated vacua, the energy density of the vacuum is
determined to be the following negative value:
\begin{equation}
 {\cal E}_{\rm temp} =  \fr{\lambda^3}{27\alpha^2} <0.
\end{equation}
Finally, we discuss the symmetry breaking pattern.
Within the Ansatz \eqref{181121.1258}, the broken symmetries are the
 temporal translation $P^0$, the Lorentz boost $M^{0m}$, the global
$U(1)$ transformation, and the shift transformation. 
However, the simultaneous transformation of the temporal translation 
and the global $U(1)$ transformation remains unbroken.
Thus, the symmetry breaking pattern is 
$ISO(3,1)\times U(1)\times S\to ISO(2,1)\times [U(1)\times {\cal P}^0]_{\rm diag.}$.

The SUSY variation of the fermion in the vacuum reads
\begin{equation}
  \delta \psi_{\alpha} = \sr{2}
i(\bar{\sigma}^0)_{\alpha \dot{\alpha}}
  \bar{\xi}^{\dot{\alpha}} \dot{\varphi}.
\end{equation}
We find again that SUSY is spontaneously broken in the
temporally modulated vacuum and in this case, the energy density is
not positive but negative.

\subsection{Lightlike modulated vacua}
For the lightlike modulation, we assume the following Ansatz:
\begin{equation}
  \vev{\varphi} = \varphi_0 e^{i\omega(x^0+x^1)}, \quad
  \vev{\der_2 \varphi} = \vev{\der_3 \varphi}
= \vev{ \psi_\alpha} 
= \vev{\b\psi_{\d\alpha}} 
= \vev{F} = \vev{\b{F}} = 0.
\end{equation}
The Ansatz implies 
$\der^m\varphi\der_m\varphi=\der^m\varphi\der_m\b\varphi=0$.
The conditions \eqref{181120.2101} and \eqref{181120.2128} are
satisfied if we demand that ${\bf L}^{10}=0$. 
Note that this condition is too strong, but it is a sufficient
condition for the lightlike modulated vacua. 
Under the condition ${\bf L}^{10}=0$, the energy-extremum condition
implies 
\begin{equation}
 \pd{{\cal L}}{\der_1 \varphi} = -k \der_1 \b\varphi = 0,
\quad
 \pd{{\cal L}}{\der_1 \b\varphi} = -k \der_1 \varphi = 0.
\end{equation}
These equations are satisfied if 
\begin{equation}
 k=0.
\end{equation}
Since $k=0$, the condition ${\bf L}^{10}=0$ leads to
$4\lambda|\d\varphi|^2=0$. 
Therefore, the parameter $\lambda$ vanishes:
\begin{equation}
 \lambda=0.
\end{equation}
With the conditions, $k=\lambda=0$, the EOM is automatically
satisfied. 
We can now calculate the vacuum energy and it vanishes identically
\begin{equation}
 {\cal E}_{\rm LL} = 0,
\end{equation}
since $k=\lambda=0$ and
$\der^m\varphi\der_m\varphi=\der^m\varphi\der_m\b\varphi=0$.

The SUSY variation of the fermion in the vacuum reads
\begin{align}
  \delta \psi_{\alpha} = \sr{2}i (\bar{\sigma}^+)_{\alpha \dot{\alpha}}
  \bar{\xi}^{\dot{\alpha}} \partial_+ \varphi,
\end{align}
where $\sigma^+=\sigma^0+\sigma^1$, $x^{+}=x^0+x^1$, and
$\partial_+=\frac{\partial}{\partial x^+}$.
Thus, SUSY is spontaneously broken in the lightlike modulated vacuum.

\section{Fluctuations around the modulated vacua}\label{sec:fluc}
In this section, we consider fluctuations of both the complex scalar
field and the fermion around the modulated vacua. 
In the previous section, we have studied the modulated 
vacua which are configurations satisfying the EOM and the
energy-extremum condition. 
Here, we discuss the local stability of the modulated vacua by
calculating the quadratic fluctuations of the dynamical fields about
the modulated vacua. 
First, we review the bosonic fluctuations in the modulated vacua
\cite{Gudnason:2018bqb}. 
Second, we consider the fermionic fluctuations.
In the modulated vacua, the fermion becomes a Goldstino
since SUSY is spontaneously broken.
We will see that the Goldstino becomes a ghost if the vacuum energy is 
negative in the modulated vacua. 

\subsection{Fluctuation of the complex scalar field}
Here, we recapitulate the fluctuation of the complex scalar field
and its stability~\cite{Gudnason:2018bqb}.
The fluctuation of the complex scalar field $\phi$ is characterized by
the value of the complex scalar field around the VEV $\vev{\varphi}$,
\begin{equation}
  \varphi \to \vev{\varphi} + \phi.
\end{equation}
In the previous section, the vacua have been characterized by 
the energy-extremum conditions.
In order to find physical vacua, we should consider the stability of
the vacua. 
The local stability of the vacua can be seen from the stability of the
fluctuation spectrum at the second order. 
Hence, we expand the energy density as follows:
\begin{align}
&  {\cal H} (\der_m \varphi, \der_m \b\varphi) \non
& = {\cal H}|_0 
+ \left.\pd{{\cal H}}{\der_m \varphi}\right|_0 \der_m \phi
+ \left.\pd{{\cal H}}{\der_m \b\varphi}\right|_0 \der_m \b\phi
+ \fr{1}{2}\mtx{\der_m \b\phi & \der_m \phi} 
{\bf M}^{mn}|_0
\mtx{\der_n \phi \\ \der_n \b\phi}
+\cdots.
\end{align}
Here, the symbol $|_0$ denotes the value at the vacuum,
the ellipses $\cdots$ mean the terms at the third order of the
fluctuation field or higher. 
The matrices ${\bf M}^{mn}$ are the second-order derivatives of the
energy density, defined in \er{eq:Mmn_def}.
Note that ${\bf M}^{\dg 00} = {\bf M}^{00}$, 
${\bf M}^{\dg 0i} = {\bf M}^{i0}$ and
${\bf M}^{\dg ij} = {\bf M}^{ji}$.
In the modulated vacua, the energy-extremum condition implies
\begin{equation}
  \left.\pd{{\cal H}}{\der_m \varphi}\right|_0 
= \left.\pd{{\cal H}}{\der_m \b\varphi}\right|_0 = 0.
\end{equation}
Thus, the energy density can be rewritten as
\begin{equation}
 {\cal H} (\der_m \varphi, \der_m \b\varphi)
= {\cal H}|_0 
+\fr{1}{2}\mtx{\der_m \b\phi & \der_m \phi} 
{\bf M}^{mn}|_0
\mtx{\der_n\phi \\ \der_n\b\phi}
+\cdots.
\end{equation}
The local stability depends on the eigenvalues of ${\bf M}^{mn}$.
If all the eigenvalues are non-negative, the vacua are locally
stable. 

The dynamics of the fluctuations is determined by the effective
Lagrangian for the fluctuation fields, which is found by expanding the
original Lagrangian around the vacua to second order:  
\begin{equation}
 {\cal L}
  = {\cal L}|_0 
+ \fr{1}{2}
\mtx{\der_m \b\phi &\der_m \phi} {\bf L}^{mn}|_0
\mtx{\der_n\phi \\ \der_n\b\phi}+\cdots.
\end{equation}
Note, that we have used that the first order variation vanishes by the
EOM: 
$\pd{{\cal L}}{\der_m \varphi}|_0 = \pd{{\cal L}}{\der_m \b\varphi}|_0 =0$.
In the following, we will consider the fluctuation Lagrangian's
corresponding stability in each of the cases of spatially, temporally,
and lightlike modulated vacua in turn.

\subsubsection{Spatially modulated vacua}
For the spatially modulated vacua, the components of the
${\bf M}^{mn}$ are 
\begin{align}
 {\bf M}^{00} &= -{\bf M}^{22} = -{\bf M}^{33} =
\mtx{
k  - \alpha |\del_1 \varphi|^4
& 
- 2 (\del_1 \varphi)^2 (\lambda + \alpha |\del_1 \varphi|^2)
\\
- 2 (\del_1 \bar{\varphi})^2 (\lambda + \alpha |\del_1 \varphi|^2)
& 
k  - \alpha  |\del_1 \varphi|^4
}, \\
 {\bf M}^{11} &=
\mtx{
 k - 4 \lambda  |\del_1 \varphi|^2 - 9 \alpha |\del_1 \varphi|^4
& 
-2 (\del_1 \varphi)^2 
(
\lambda  + 3 \alpha  |\del_1 \varphi|^2
)
\\
-2 (\del_1 \bar{\varphi})^2 
(
\lambda  + 3 \alpha  |\del_1 \varphi|^2
)
& 
 k - 4 \lambda |\del_1 \varphi|^2 - 9 \alpha  |\del_1 \varphi|^4
},  
\end{align} 
whereas the remaining components of ${\bf M}^{mn}$ vanish.
Since the generalized mass matrix $\mathcal{M}$ is totally block
diagonal, the local stability of the modulated vacua is determined
from the spectrum of eigenvalues of the matrices ${\bf M}^{mn}$.
For the matrices ${\bf M}^{00}$, ${\bf M}^{22}$ and ${\bf M}^{33}$,
the eigenvalues $A_1, A_2$ are
\begin{equation}
  A_1 = 0, \quad A_2 = 
\fr{12\alpha k -4\alpha \lambda (\lambda +\sr{\lambda^2 + 3k \alpha})}
{9\alpha^2},
\end{equation}
whereas the eigenvalues $B_1, B_2$ of the matrix ${\bf M}^{11}$ are
\begin{equation}
 B_1 = 0, \quad B_2 =
 -\fr{4}{3\alpha} (\lambda^2 + 3\alpha k +\lambda \sr{\lambda^2 +3\alpha k}).
\end{equation}
The zero eigenvalue of ${\bf M}^{11}$ corresponds to the generalized
NG mode due to the broken translational symmetry. 
The zero eigenvalues of ${\bf M}^{22}$ and ${\bf M}^{33}$ 
originate from the rotational symmetry $SO(3) \subset SO(1,3)$ 
of the original Lagrangian.
In the region where $\alpha<0$, $\lambda>0$ and
$\lambda^2+3\alpha k>0$, the nonzero eigenvalues are positive.
Since there are no negative eigenvalues, the modulated vacua are
locally stable. 

The fluctuations in the Lagrangian can also be computed.
The eigenvectors corresponding to the zero eigenvalues of 
the matrix ${\bf M}^{00}$ are nondynamical up to the
second order of the fluctuations.
Explicitly, the Lagrangian is 
\begin{equation}
 {\cal L} = {\cal L}|_0 
+ \fr{1}{2} \mtx{\d{\b\phi} & \d\phi} {\bf M}^{00}|_0
 \mtx{\d\phi \\ \d{\b\phi}}
+ \cdots.
\end{equation}
Here, we have used ${\bf M}^{00}|_0 = {\bf L}^{00}|_0$ in the
spatially modulated vacua. 
Therefore, the eigenvector associated with the zero eigenvalue 
is not dynamical.
Since the zero modes of the generalized mass matrix and those of the 
matrices $\mathbf{L}^{mn}$ coincide in the spatially modulated vacua
in general, the canonical kinetic terms for the generalized NG modes
disappear in those vacua \cite{Nitta:2017mgk}. 
It has been shown that this, however, is not the case for temporal and
lightlike modulated vacua \cite{Gudnason:2018bqb} as we will see 
below.

\subsubsection{Temporally modulated vacua}
For the temporally modulated vacua, the matrices ${\bf M}^{mn}$ are
calculated as 
\begin{align}
{\bf M}^{00} &= 
\mtx{
k + 12\lambda |\d\varphi|^2 -45 \alpha |\d\varphi|^4 
& 6\d\varphi^2 (\lambda -5\alpha |\d\varphi|^2) \\
6\d{\b\varphi}^2 (\lambda -5\alpha |\d{\b\varphi}|^2)
& k + 12\lambda |\d\varphi|^2 -45 \alpha |\d\varphi|^4 
},\\
{\bf M}^{11} &= {\bf M}^{22} = {\bf M}^{33} = 
\mtx{
k+ 3\alpha |\d\varphi|^4
& 2\d\varphi^2(-\lambda + 3\alpha|\d\varphi|^2) \\
2\d{\b\varphi}^2(-\lambda + 3\alpha|\d{\b\varphi}|^2)
& k+ 3\alpha |\d\varphi|^4},
\end{align}
and the others vanish.
Again, the generalized mass matrix $\mathcal{M}$ is block diagonal.
The eigenvalues of the matrices can be calculated as follows.
For the matrix ${\bf M}^{00}$, 
the eigenvalues $A_1$ and $A_2$ are
\begin{equation}
A_1 = 0, \quad
A_2=  -\fr{8\lambda^2}{3\alpha}.
\end{equation}
The nonzero eigenvalue, $A_2$, is positive because $\alpha$ must be
negative in temporally modulated vacua (see Sec.~\ref{sec:tmv}). 
The zero mode, $A_1$, is interpreted as a generalized NG mode for the
temporally modulated vacua. 

For the matrices ${\bf M}^{11}$, ${\bf M}^{22}$, and ${\bf M}^{33}$,
the eigenvalues $B_1$ and $B_2$ vanish
\begin{equation}
 B_1 = B_2 = 0.
\end{equation}
These zeromodes are expected to be accidental, since translational
invariance along the spatial directions is not broken in the
temporally modulated vacua. 
Hence, there are no unstable modes in the bosonic fluctuation spectrum 
up to the second order in the fluctuations. 

The dynamics of the bosonic fluctuations is determined by the
effective Lagrangian for the fluctuation. 
Since the matrix ${\bf L}^{00}$ vanishes in the temporally modulated
vacua, the fluctuations are not dynamical up to second order in
derivatives. 
However, the fluctuations have a nonvanishing spatial dispersion
relation because there is a nonzero eigenvalue in
${\bf L}^{11} = {\bf L}^{22} = {\bf L}^{33}$:
\begin{equation}
{\bf L}^{11} = 
\mtx{
-k +\alpha |\d\varphi|^4 
&-2\d\varphi^2 (\lambda - 2\alpha |\d\varphi|^2)\\
-2\d{\b\varphi}^2 (\lambda - 2\alpha |\d\varphi|^2)
&-k +\alpha |\d\varphi|^4} .
\end{equation}
The eigenvalues $s_1$ and $s_2$ are
\begin{equation}
 s_1 = 0, \quad
 s_2= \fr{8\lambda^2}{9\alpha}.
\end{equation}
Since $s_2$ is positive, there are no unstable modes in the 
spatial-derivative sector.

\subsubsection{Lightlike modulated vacua}
For the lightlike modulated vacua, the matrices ${\bf M}^{mn}$ are
\begin{equation}
 {\bf M}^{00}  = {\bf M}^{11} = -{\bf M}^{01} = -{\bf M}^{10} =
\mtx{
- 16\alpha |\d\varphi|^4
&
 -8\alpha\d\varphi^2|\d\varphi|^2
\\
 -8\alpha\d{\b\varphi}^2|\d\varphi|^2
&
-16\alpha |\d\varphi|^4
},
\end{equation}
and the remaining matrices, including ${\bf M}^{22}$ and
${\bf M}^{33}$, vanish.
Here, we have already used the condition $k=\lambda=0$ which is
necessary for the existence of the lightlike modulated vacua.
Since the temporal and spatial modulations are mixed in the lightlike
modulated vacua, it is convenient to switch to light-cone coordinates
\begin{equation}
\mtx{x^+ \\ x^-} = 
\mtx{1 & 1\\ 1 & -1} \mtx{x^0 \\ x^1}. 
\end{equation}
In these coordinates, the matrices ${\bf M}^{mn}$ are simply expressed
as 
\begin{align}
 {\bf M}^{--}
&= {\bf M}^{00} -{\bf M}^{01} -{\bf M}^{10}+ {\bf M}^{11}
= 4
\mtx{
  - 16\alpha |\d\varphi|^4
  & -8\alpha\d\varphi^2|\d\varphi|^2\\
  -8\alpha\d{\b\varphi}^2|\d\varphi|^2
  &-16\alpha |\d\varphi|^4}, \\
 {\bf M}^{++} &=  {\bf M}^{+-} =  {\bf M}^{-+} = 0.
\end{align}
The eigenvalues $A_1$, $A_2$ of ${\bf M}^{--}$ are 
\begin{equation}
 A_1 = -32 \alpha \omega^4 |\varphi_0|^4, \quad
 A_2 = -96 \alpha \omega^4 |\varphi_0|^4.
\end{equation}
In order for all the eigenvalues to be positive, we require that 
\begin{equation}
 \alpha < 0.
\end{equation}
With this condition, the lightlike modulated vacua are locally stable.

Up to the second order in the fluctuation, there is no dynamics of the
fluctuations in the lightlike modulated vacua.
This is because all the matrices ${\bf L}^{mn}$ vanish in the
lightlike modulated vacua. 
For example, the matrices along $x^0 $ or $x^1$ directions are
\begin{equation}
 {\bf L}^{00} = {\bf L}^{11} = -{\bf L}^{01} = -{\bf L}^{10} = 
2\lambda 
\mtx{
|\d\varphi|^2 
&
{\d\varphi}^2
\\
\d{\b\varphi}^2
&
|\d\varphi|^2
} =0.
\end{equation}
Note, that the fluctuations may become dynamical due to higher-order
terms than the quadratic ones.

\subsection{Fluctuation of the fermion}
Since we consider a supersymmetric theory, we should consider the 
stability in the fermionic sector in the spatially, temporally, and
lightlike modulated vacua as well.
We will thus consider each case in turn in the following.

\subsubsection{General arguments}
We will now discuss the general arguments for the fluctuations of
the fermion in spatially, temporally, or lightlike modulated vacua. 
The first observation is that the fermion becomes a Goldstino in the
modulated vacua. 
This is due to the nonvanishing SUSY transformation of the fermion in
the modulated vacua: 
\begin{equation}
 \vev{\delta \psi_\alpha} 
=\sr{2}
i(\sigma^m)_{\alpha\d\beta}\b\xi^{\d\beta} 
\vev{\der_m \varphi}\neq 0.
\end{equation}
Here, $\b\xi^{\d\beta}$ is a SUSY transformation parameter,
and we have used that $\vev{F} = 0$ in the modulated vacua. 
The nonvanishing SUSY transformation of the fermion implies the
existence of the Goldstino.

The kinetic term for the Goldstino can be found by expanding the
Lagrangian around the modulated vacua, which up to the second order in
the fluctuations reads 
\begin{align}
\mathcal{L}_{\rm f.kin} =& 
\int d^4\theta \(
k \Phi\b\Phi
+\frac{1}{16} 
\(
\lambda + \alpha \partial_m \Phi \partial^m\b\Phi
\right) (D\Phi)^2 (\bar{D}\b\Phi)^2
\) \non 
=& 
- ik\b\psi\b\sigma^m \der_m \psi 
 + (\lambda + \alpha |\partial_m \varphi|^2) \Omega \non
& 
- i \alpha (\partial^m \psi \sigma^p \bar{\psi}) (\partial_m
 \bar{\varphi}) (\partial_n \varphi)^2 (\partial_p \bar{\varphi})
- i \alpha (\partial^m \bar{\psi} \bar{\sigma}^p \psi) (\partial_m
 \varphi) (\partial_n \bar{\varphi})^2 (\partial_p \varphi)
 \non
& + \cdots.
\label{181213.1116}
\end{align}
Here, the ellipses $\cdots$ denote higher order terms in the
fluctuations and are hence irrelevant for the kinetic term of the
Goldstino. $\Omega$ is defined by  
\begin{align}
\Omega 
&\equiv
- \frac{i}{2} (\psi \sigma^m \bar{\sigma}^n \sigma^p \partial_p
 \bar{\psi}) (\partial_m \varphi \partial_n \bar{\varphi})
+ \frac{i}{2} (\partial_p \psi \sigma^p \bar{\sigma}^m \sigma^n
  \bar{\psi}) (\partial_m \varphi \partial_n \bar{\varphi}) \non
&\phantom{=\ }
+ i (\psi \sigma^m \partial^n \bar{\psi}) (\partial_m \varphi
 \partial_n \bar{\varphi})
- i (\partial^m \psi \sigma^n \bar{\psi}) (\partial_m \varphi \partial_n
 \bar{\varphi}).
\end{align}
In this expansion, we have used the fact that the fermion appears only
at second order in the auxiliary field: $F = 0 + \mathcal{O} (\psi^2)$
on the canonical branch \cite{Nitta:2017yuf}.

In the following, we will explicitly show the stability of the
fluctuation of the Goldstino in the modulated vacua up to second
order. 
The stability of the Goldstino in the modulated vacua depends on the
sign of its kinetic term.
In the Lagrangian, the sign of the time-derivative term
$i\b\psi\b\sigma^0\der_0\psi$ of the Goldstino is given by that of 
the parameter $k$, which we assume to be positive: $k \geq 0$.
However, the sign of the kinetic term can be altered by the presence
of the VEV of the complex scalar field in the modulated vacua.
If the kinetic term has the correct (wrong) sign, the fluctuation of
the Goldstino is stable (unstable). 
For the spatially modulated vacua, there are metastable, degenerate,
and unstable vacua. 
For the temporally modulated vacua, the Goldstino becomes a ghost, and
the vacua become unstable. 
For the lightlike modulated vacua, the Goldstino is not dynamical.
In the following, we will study the stability explicitly for each case
in turn.

\subsubsection{Spatially modulated vacua}
For the spatially modulated vacua, 
the sign of the kinetic term depends on the model parameters as well
as the vacuum solution as follows:
\begin{align}
\mathcal{L}_{\rm f. kin} 
&=
 i \(-k +\lambda |\der_1 \varphi|^2 + \alpha |\der_1 \varphi|^4\)
\b\psi_{\d\alpha} (\b\sigma^0)^{\d\alpha \beta} \der_0 \psi_\alpha
\non
&= 
 -i \fr{{\cal E}_{\rm sp}}{|\vev{\der_1 \varphi}|^2}
\b\psi_{\d\alpha} (\b\sigma^0)^{\d\alpha \beta} \der_0 \psi_\alpha.
\end{align}
The sign of the kinetic term is thus related to that of the vacuum
energy density \eqref{181011.1320}. 
We can see that if the energy density is positive (negative), the 
Goldstino has the correct (wrong) sign for its kinetic term.
This property was clarified in Ref.~\cite{Nitta:2017yuf}.
In Sec.~\ref{gg} we will see that this relation is consistent with the 
analysis using the SUSY algebra.

\subsubsection{Temporally modulated vacua}
For the temporally modulated vacua, the kinetic term of the Goldstino
becomes 
\begin{align}
\mathcal{L}_{\text{f.kin}} &=
i 
(-k - 3 \omega^2 |\varphi_0|^2 \lambda + 5 \alpha \omega^4 |\varphi_0|^4)
\bar{\psi}_{\d\alpha} (\bar{\sigma}^0)^{\d\alpha\alpha} 
\partial_0 \psi_\alpha + \cdots
\non
&= 
-i \fr{{\cal E}_{\rm temp}}{|\vev{\d\varphi}|^2}
\bar{\psi}_{\d\alpha} (\bar{\sigma}^0)^{\d\alpha\alpha} 
\partial_0 \psi_\alpha,
\end{align}
where we have used the relations 
$k = - \frac{\lambda^2}{3 \alpha}$, 
$\omega^2 |\varphi_0|^2 = \frac{\lambda}{3 \alpha}$,
and ${\cal E}_{\rm temp}=\fr{\lambda^3}{27\alpha^2}$. 
Since the energy density is negative, ${\cal E}_{\rm temp}<0$, we can
conclude that the Goldstino is a ghost Goldstino in the temporally
modulated vacua.

\subsubsection{Lightlike modulated vacua}
\label{flm}
For the lightlike modulated vacua, however, the quadratic kinetic
term in \er{181213.1116} vanishes.
This can be shown as follows.
The existence of the lightlike modulated vacua requires
$k=\lambda=0$. 
Therefore, the Lagrangian up to second order in fermionic fluctuations
becomes 
\begin{equation}
{\cal L}_{\rm f. kin}
=   \alpha |\partial_m \varphi|^2 \Omega 
- i \alpha (\partial^m \psi \sigma^p \bar{\psi}) (\partial_m
 \bar{\varphi}) (\partial_n \varphi)^2 \partial_p \bar{\varphi}
- i \alpha (\partial^m \bar{\psi} \bar{\sigma}^p \psi) (\partial_m
\varphi) (\partial_n \bar{\varphi})^2 \partial_p \varphi
+\cdots.
\end{equation}
However, the VEVs of $\der^m \varphi\der_m \varphi$ 
and $\der^m \varphi\der_m \b\varphi$ vanish in the
lightlike modulated vacua
\begin{equation}
\vev{ \der^m \varphi\der_m \varphi}  =
\vev{ \der^m \varphi\der_m \b\varphi} = 0,
\end{equation}
and thus the kinetic term of the Goldstino vanishes too.
Thus, the Goldstino is not dynamical.
This property is consistent with the fact that the vacuum energy
density vanishes in the lightlike modulated vacua.

\section{Vacuum energy density vs stability of Goldstino}
\label{gg}
In this section, we will derive the relation between the sign of the
kinetic term of the Goldstino and that of the vacuum energy density in
the modulated vacua. 
In the previous sections, we have used a specific model for the
modulated vacua, whereas the relation that we will demonstrate in this
section is model independent as it is based entirely on the SUSY
algebra and the preserved symmetries of the model (and corresponding
modulated vacuum) at hand.

Since the fermion becomes the Goldstino, the dynamics of the
fluctuation of the fermion can be kinematically discussed by using the
SUSY algebra {\it i.e.}~the relation between SUSY and the
Hamiltonian,
\begin{equation}
H = \df{1}{4} 
(Q_1 \b{Q}_{\d{1}} + Q_2 \b{Q}_{\d{2}} +
\b{Q}_{\d{1}} Q_1 + \b{Q}_{\d{2}}Q_2)
=: \df{1}{4}\sum_{\rm \alpha:spinors} \b{Q}_{\alpha} Q_\alpha.
\end{equation}
By considering the vacuum expectation value of both sides, one can
show that there is a ghost Goldstino when the vacuum energy is
negative.
We thus apply this to the discussion of the modulated vacua.

However, the translational generators along spatial or temporal
directions may not be well-defined operators in the modulated vacua.
This problem is caused by the divergence of the spatial integration
of a charge operator.
Therefore, we should discuss the relation between SUSY and the
Hamiltonian in a system with finite (but large) volume $V$ 
with periodic boundary conditions to preserve translational invariance
along spatial directions. 
The following discussion is similar to the one in
Ref.~\cite{Weinberg:2000cr}.  

\subsection{Temporally modulated vacua}
\label{ggtm}
We will now show that the Goldstino is a ghost in the temporally
modulated vacua in the case where the model has negative vacuum
energy. 
First, we consider the vacuum expectation value of the relation
between the Hamiltonian and the supercharges due to the SUSY algebra. 
For the temporally modulated vacua, we should discuss the relation in
a finite volume. 
The vacuum expectation values read 
\begin{equation}
 \bra{\rm vac,\, box} H \ket{\rm vac,\, box}
 = 
\df{1}{4}\sum_{\rm \alpha:spinors} 
\bra{\rm vac,\, box}
\b{Q}_{\alpha} Q_\alpha
\ket{\rm vac,\, box},
\label{180820.0053}
\end{equation}
where the ket $\ket{\rm vac, \, box}$ denotes the vacuum state of the
system in a box with the above discussed periodic boundary conditions.
We will show that the right-hand side of \er{180820.0053} is the norm
of the Goldstino one-particle state. 
We expand the right-hand side by inserting multiparticle 
states normalized by the finite volume $\ket{X, \, {\rm box}}$ 
as follows:
\begin{equation}
 \bra{\rm vac, \, box} H \ket{\rm vac, \, box}
 = 
\df{1}{4}\sum_{ X, \, \alpha:{\rm spinors}} 
|\bra{X, \, \rm box} Q_\alpha
\ket{\rm vac,\, box}|^2.
\label{180819.2321}
\end{equation}
Now, we consider a case where the energy density has the vacuum
expectation value ${\cal E}$: 
\begin{equation}
  \bra{\rm vac, \, box} H \ket{\rm vac, \, box}
  = V {\cal E}.
\end{equation}
Relating the ket in a finite system to that of an infinite system, we
get 
\begin{equation}
\ket{X, \, {\rm box}} 
\to \(\sr{\df{(2\pi)^3}{V}}\)^{N_X} \ket{X}, 
\end{equation}
where $N_X$ denotes the number of particles in the state $X$.
By this replacement, \er{180819.2321} can be rewritten as
\begin{equation}
V{\cal E} = 
\df{1}{4}
\sum_{ X, \, \alpha:{\rm spinors}} 
\df{(2\pi)^{3N_X}}{V^{N_X}}
|\bra{X} Q_\alpha
\ket{\rm vac}|^2. 
\label{180820.0112}
\end{equation}
We argue that only the zero-momentum state in $\ket{X,\,\rm box}$
contributes on the right-hand side.
This is because $Q_\alpha \ket{\rm vac, \, box}$ belongs to the same
eigenstate of the three-momentum as $\ket{\rm vac, \, box}$, since
$[Q,P_i]=0$ for $i=1,2,3$ holds. 
Therefore, only the eigenstates with zero three-momentum 
$\ket{X(\bs{p}=\bs{0})}$ in $\ket{X}$ 
contribute to the right-hand side of \er{180819.2321}.
Therefore, 
$Q_{\alpha} = \int \! d^3 x \ S^{m=0}_{\alpha} (x^0, \mathbf{x})$
can be reduced into the product of the volume
$V$ and the supercurrent $S^m_\alpha (x^0,\bs{x})$ at $\bs{x}=\bs{0}$: 
\begin{align}
 |\bra{X} Q_\alpha
\ket{\rm vac}|^2
&=
 \left|\int d^3x\;\bra{X  \,(\bs{p}=\bs{0})} S^0_\alpha(x^0,\bs{x})
\ket{\rm vac}\right|^2 \non
&=
  \left|\int d^3x\;\bra{X(\bs{p}=\bs{0})}
  e^{i\bs{P}\cdot \bs{x}} S^{0}_\alpha(x^0, \bs{x}=\bs{0})
  e^{-i\bs{P}\cdot \bs{x}}\ket{\rm vac}\right|^2 \non
&= \left|\int d^3x\;\bra{X(\bs{p}=\bs{0})}
  S^{0}_\alpha(x^0, \bs{x}=\bs{0})\ket{\rm vac}\right|^2 \non
&= V^2 |\bra{X(\bs{p}=\bs{0})} S^{0}_\alpha(x^0, \bs{x}=\bs{0})
\ket{\rm vac}|^2,
\label{180820.0111}
\end{align}
where we have used that both the vacuum and the state $X$ are
zero-momentum states and hence the integral, finally, is independent
of $\bs{x}$ and hence proportional to the volume.
By using the above equation, \er{180820.0112} can be written as 
\begin{equation}
{\cal E}
 = 
\df{1}{4}
\sum_{ X , \, \alpha:{\rm spinors}} 
\df{(2\pi)^{3N_X}}{V^{N_X-1}}
 |\bra{X \,  (\bs{p}=\bs{0})} S^{0}_\alpha(x^0, \bs{0})
\ket{\rm vac}|^2. 
\label{180820.0127}
\end{equation}
In the limit $V\to\infty$, the dominant but finite contributions to
the right-hand side come from the states with $N_X =1$. 
Note, that the contribution from the zero-particle state should vanish
since such a contribution diverges in the limit $V\to\infty$ while the
left-hand side is finite. 
For the states with $N_X=1$, we find the relation 
\begin{equation}
{\cal E}
 = 
\df{1}{4}
\sum_{ X , \, \alpha:{\rm spinors}} 
(2\pi)^{3}
 |\bra{X \,  (\bs{p}=\bs{0},\, N_X =1)} S^{0}_\alpha(x^0, \bs{0})
\ket{\rm vac}|^2. 
\label{180820.0128}
\end{equation}
If the vacuum energy density is nonzero ${\cal E}\neq 0$, 
the state $ S^{0}_\alpha(x^0, \bs{0})\ket{\rm vac}$
carries one particle state with $\bs{p}= \bs{0}$,
which is identified as the Goldstino.
Further, \er{180820.0128} can be seen as a norm of the 
one particle state $ S^{0}_\alpha(x^0, \bs{0})\ket{\rm vac}$.
Therefore, the norm of the Goldstino is negative if 
the vacuum energy density is negative.

\subsection{Spatially modulated vacua}
\label{ggsm}
We will now show the relation between the negative vacuum energy
and the ghost Goldstino in the spatially modulated vacua.
In the spatially modulated vacua, the discussion is almost 
the same as in Sec.~\ref{ggtm}, except for the fact that the 
spatial translation ($P_1$) is broken, in spatially modulated vacua.

If we assume the phase of the vacuum expectation value of the 
complex scalar field is modulated as 
$\vev{\varphi} =\varphi_0 e^{ic x^1 }$, 
we can argue that the simultaneous transformation with $P_1$ 
and the global $U(1)$, is preserved.
Here, we assume that $A \varphi =  q \varphi$, where $A$ is
the Hermitian generator of the $U(1)$ transformation, 
and $q$ is the charge of the complex scalar field.
The unbroken operator is then given by 
\begin{equation}
 P_1^{\rm S} := P_1 - \fr{c}{q} A,
\end{equation}
where $P_1$ is the Hermitian generator of the translation along the
$x^1$ direction: $P_1 \varphi = -i \der_1 \varphi$. 
Thus, we should try to use $P^{\rm S}_1$ instead of $P_1$.
The only fact we need is that the $U(1)$ generator $A$ commutes with
the SUSY charge $Q_\alpha$,
\begin{equation}
[A,Q_\alpha]=0.
\end{equation}
The state $X$ that contributes in \er{180820.0111} is the one with
$p_2 = p_3 =0$, but finite momentum and finite $U(1)$ charge $q$: 
$P_1 \ket{X} =  c \ket{X}$ and 
$\fr{c}{q}A \ket{X} = c \ket{X}$, respectively.
This is because the conserved quantity is $P_1^S$, which corresponds
to a translation and simultaneous local $U(1)$ transformation.
Therefore, the relation between the supercurrent and the 
vacuum energy becomes
\begin{align}
 |\bra{X} &Q_\alpha
\ket{\rm vac}|^2
= \left|\int d^3 \bs{x}\bra{X(p_2 = p_3= 0, p_1 = c)}
S^0_\alpha(x^0,\bs{x}) \ket{\rm vac}\right|^2 \non
&= \left|\int d^3x\;\bra{X(p_2 = p_3= 0, p_1 = c)}
  e^{i\bs{x} \cdot \bs{P}} e^{\frac{-i x^1 c A}{q}}  e^{\frac{i x^1 c A}{q}}
  S^{0}_\alpha(x^0, \bs{x}=\bs{0})
  e^{-i\bs{x} \cdot \bs{P}}\ket{\rm vac}\right|^2 \non
&= \left|\int d^3x\;\bra{X(p_2 = p_3= 0, p_1 = c)}
  e^{i\bs{x} \cdot \bs{P}} e^{\frac{-i x^1 c A}{q}}  
  S^{0}_\alpha(x^0, \bs{x}=\bs{0}) e^{\frac{i x^1 c A}{q}}
  e^{-i\bs{x} \cdot \bs{P}}\ket{\rm vac}\right|^2 \non
&=\left|\int d^3x\;\bra{X(p_2 = p_3= 0, p_1 = c)}
  e^{i\bs{x} \cdot \bs{P}^{\rm S}}
  S^{0}_\alpha(x^0, \bs{x}=\bs{0}) 
  e^{-i\bs{x} \cdot \bs{P}^{\rm S}}\ket{\rm vac}\right|^2 \non
&=\left|\int d^3x\;\bra{X(p_2 = p_3= 0, p_1 =c)}
  S^{0}_\alpha(x^0, \bs{x}=\bs{0})\ket{\rm vac}\right|^2 \non 
& = V^2 
 |\bra{X(p_2 = p_3= 0,p_1 = c)} S^{0}_\alpha(x^0, \bs{x}=\bs{0})
\ket{\rm vac}|^2,
\end{align}
where we have inserted a $U(1)$ transformation together with its
inverse on the left-hand side of the supercurrent and commuted the
inverse transformation to the other side of the latter.
The resulting vector $\bs{P}^{\rm S} = (P_1^{\rm S}, P_2, P_3)$ is a
set of operators for the unbroken symmetries.

By the same argument as in the case of the temporally modulated vacua, 
we conclude that the one-particle state
$|\bra{X(p_2 = p_3= 0,p_1 = c,N_X =1)}S^{0}_\alpha(x^0,\bs{0})\ket{\rm vac}|^2$ 
becomes a ghost if the vacuum energy is negative.

\subsection{Lightlike modulated vacua}
For the lightlike modulated vacua, it will be convenient to use 
the light-cone coordinates. 
The symmetry breaking pattern in this case is
$U(1) \times {\cal P}^0\times {\cal P}^1\to [U(1) \times {\cal P}^{\pm}]_{\rm diag} \times {\cal P}^{\mp}$,
where the ${\cal P}^{\pm}$ represents the translational symmetry group
along the light-cone directions
$x^\pm = x^0 \pm x^1$.
For the VEV $\vev{\varphi} = \varphi_0 e^{i\omega (x^0 + x^1)}$,
the symmetry breaking pattern becomes
$U(1) \times {\cal P}^0\times {\cal P}^1\to [U(1) \times {\cal P}^+]_{\rm diag} \times {\cal P}^-$.
Note, that we define the Hermitian generator of the translational 
group ${\cal P}^+$ and ${\cal P}^-$ as
\begin{equation}
 P_+ := P_0 + P_1, 
\quad
 P_- := P_0 - P_1,
\end{equation}
respectively.
We again assume that the $U(1)$ charge of the complex scalar field
is $q$: $A \varphi = q \varphi$.
With this assumption, the unbroken generator corresponding to the
unbroken group $[U(1) \times {\cal P}^+]_{\rm diag}$ 
can explicitly be written as 
\begin{equation}
P^{\rm L}_+ := P_+ - \fr{2\omega}{q} A,
\end{equation}
where we have used $P_+ \varphi = 2\omega \varphi$.
Thus, the unbroken translational operator along the $x^1$ direction
$P_1^{\rm L}$ can be written in terms of unbroken generators as
\begin{equation}
  P_1^{\rm L}  = \fr{1}{2}(P^{\rm L}_+ - P_-) = P_1 -  
\fr{\omega}{q} A,
\end{equation}
which is also an unbroken operator.
Since $A$ commutes with the SUSY generator, we can repeat the argument
of Sec.~\ref{ggsm} by replacing $c$ with $\omega$.
Thus, the relation between the sign of the vacuum energy density 
and the norm of the Goldstino also holds in the lightlike 
modulated vacua.
In particular, the Goldstino becomes a zero-norm state in the vacua
where the vacuum energy density vanishes, which agrees with our result
in Sec.~\ref{flm}.

\section{Summary and discussion}\label{sec:summary}
In this paper, we have explored a new spontaneous SUSY-breaking
mechanism with spatially, temporally, or lightlike  
modulated vacua.
We have used a ghost-free SUSY higher-derivative model with a chiral
superfield, which is a supersymmetric extension \cite{Nitta:2017yuf} of
the model used in Refs.~\cite{Nitta:2017mgk,Gudnason:2018bqb}.

In this model, all the spatially, temporally, or lightlike modulated
vacua are realized as the energy-extremum state and the solution to
the EOM within the Ansatz of phase modulation.
We have calculated the vacuum energy density of each of the modulated
vacua. 
For the spatially modulated vacua, 
the vacuum energy can be positive, zero, or negative,
depending on the choice of the parameter of the model.
For the temporally modulated vacua, 
the vacuum energy density is always negative in our model.
For the lightlike modulated vacua, the vacuum energy density
vanishes. 

We have then investigated the stability of the fluctuation
around the modulated vacua.
For the bosonic fluctuation given by a complex scalar field, 
there are stable and nondynamical fluctuations while there are no
unstable modes in any of the modulated vacua. 
This property coincides with the non-SUSY case \cite{Gudnason:2018bqb}.
However, for the fermionic fluctuations, there are unstable ghost
modes in the spatially or temporally modulated vacua. 
We have argued that the ghost can be related to the negative vacuum
energy density of the modulated vacuum, using the SUSY algebra.

There are several possible directions for future work.
One is to discuss the instability of the temporally modulated vacua
in supersymmetric theories.
The temporally modulated vacuum is unstable due to the negative vacuum
energy density in our SUSY model, in contrast to the non-SUSY case
\cite{Gudnason:2018bqb}, where the temporally modulated vacuum is
stable. 
It is plausible that this instability is model dependent, however,
future investigations are needed for such a conclusion.
It may also be possible that the vacuum energy density is uplifted by
higher-order terms such as $(\der\phi)^{8}$ and the Goldstino becomes
a physical fluctuation. 
As another possibility, SUSY might forbid stable temporally modulated
vacua.  
In such a case, we should discuss the instability in a more model
independent way.
The application of our model to more realistic phenomenological models 
with metastable SUSY breaking modulated vacua would be interesting. 
We have studied SUSY breaking in a higher-derivative chiral superfield
in this paper. 
An extension to a vector superfield would also be possible, since the
most general higher-derivative vector superfield action, free from 
ghosts and the auxiliary field problem, is available
\cite{Fujimori:2017kyi}. 
We will leave these questions for future work.

\subsection*{Acknowledgments}

This work is supported by the Ministry of Education,
Culture, Sports, Science (MEXT)-Supported Program for the Strategic Research Foundation at Private Universities ``Topological Science'' (Grant No.\ S1511006).
The work of M.~N.~is also supported in part by  
the Japan Society for the Promotion of Science
(JSPS) Grant-in-Aid for Scientific Research (KAKENHI Grants
No.~16H03984 and No.~18H01217).
The work of M.~N.~and S.~B.~G.~is also supported in part by a Grant-in-Aid for 
Scientific Research on Innovative Areas ``Topological Materials
Science'' (KAKENHI Grant No.~15H05855) from the MEXT of Japan. 
The work of S.~S. is supported by JSPS KAKENHI Grant No.~JP17K14294.
The work of R.~Y.~is supported in part by a Grant-in-Aid for
Scientific Research on Innovative Areas ``Discrete Geometric Analysis
for Materials Design'' (KAKENHI Grant No.~17H06462) from the MEXT of
Japan.

\end{document}